# A compact acoustic calibrator for ultra-high energy neutrino detection

S. Adrián-Martínez, M. Ardid, M. Bou-Cabo, G. Larosa, C.D Llorens, J.A. Martínez-Mora

Institut d'Investigació per a la Gestió Integrada de les Zones Costaneres, Universitat Politècnica de València, C/ Paranimf 1, 46730 Gandia, València, Spain.

Corresponding autor. Tel.: +34 962849314; Fax: +34 962849309
E-mail address: siladmar@upv.es (S. Adrián-Martínez)

**ABSTRACT**

With the aim to optimize and test the method of acoustic detection of ultra-high energy neutrinos in underwater telescopes a compact acoustic transmitter array has been developed. The acoustic parametric effect is used to reproduce the acoustic signature of an ultra-high-energy neutrino interaction. Different R&D studies are presented in order to show the viability of the parametric sources technique to deal with the difficulties of the acoustic signal generation: a very directive transient bipolar signal with 'pancake' directivity. The design, construction and characterization of the prototype are described, including simulation of the propagation of an experimental signal, measured in a pool, over a distance of 1 km. Following these studies, next steps will be testing the device in situ, in underwater neutrino telescope, or from a vessel in a sea campaign.

*Keywords:*
Underwater neutrino telescopes; acoustic neutrino detection; parametric acoustic sources; underwater acoustic array sensors.

## 1. Introduction

The method of acoustic detection is a promising technique for the detection of cosmic neutrinos with energies exceeding 1 EeV. Askariyan in 1957 published the first studies in relation with the possibility of detecting ionizing particles by acoustic techniques [1]. When a neutrino interacts with a nucleus of water, a particle shower is generated. On average, 25% of the neutrino energy is deposed in a cylindrical volume of a few cm of radius and several meters of length and it produces an instantaneous, with respect to the hydrodynamic time scales, local heating of the medium. As a consequence of the temperature change and depending on the volume expansion coefficient of the medium, an expansion or contraction of the medium is induced. A pressure pulse of bipolar shape is propagated in the surrounding medium due the accelerated expansion of the heated volume [2-3] with, due to the dimensions of the source, a 'pancake' directivity. This means a flat disk emission pattern perpendicularly to the axis defined by the hadronic shower. As a reference example at 1 km distance in perpendicular direction to a $10^{20}$ eV hadronic shower, it is expected that the bipolar acoustic pulse has about 0.2 Pa (peak-to-peak) in amplitude and about 50 µs length (peak-to-peak). With respect to the directivity pattern the opening angle of the pancake is expected to be about 1º. Ultra-high energy (UHE) acoustic detection is a promising technique to extent the energy range of different experiments. In that sense, ANTARES [4] is considering the technique, using the detector as platform to test the feasibility of a large-scale acoustic neutrino detector to combine optical and acoustical detection techniques for hybrid underwater neutrino telescopes, especially considering that the optical neutrino technique needs acoustic sensors as well for positioning purposes. ANTARES contains AMADEUS [5], an acoustic test setup which is operational and taking data. It can be considered a prototype to evaluate the feasibility of the neutrino acoustic detection. Even though all AMADEUS sensors have been characterized in the lab, it would be desirable to deploy a compact calibrator that "in situ" may be able to calibrate the detection system, to train and tune the system in order to improve its performance, to test and validate the technique, and to determine the reliability of the system [6].

## 2. Previous studies

Previous research studies were done to evaluate the possibility of using the parametric acoustic sources technique to reproduce the acoustic signature of a UHE neutrino interaction.

Acoustic parametric generation was first proposed by Westervelt [7] in the 1960's. The acoustic parametric effect occurs when two intense monochromatic beams, with two close frequencies, travel together through a medium. Under these conditions, in the region of nonlinear interaction, secondary harmonics of these frequencies are produced: the sum and difference of the frequencies and the double frequencies of both beams. The main advantage of this technique is that the secondary parametric beam has the same directivity pattern as the primary beam, enabling low frequency beams (difference frequency) with high directivity (primary beam). Since the emission is made at high frequencies, it is possible to obtain narrow directional patterns using a transducer with small overall dimensions. However there are some difficulties when applying this method to a compact acoustic calibrator as the neutrino acoustic signal is a transient signal with broad frequency content with a cylindrically symmetric directivity. To deal with transient signals, theoretical and experimental studies [8] indicate that it is possible to generate a signal with 'special' modulation at larger frequency in such a way that the pulse interacts with itself while it is travelling through the medium, thereby generating the desired signal. In this case, the secondary parametric signal generated in the medium is related to the second time derivative of the envelope of the primary signal, following the equation:

$$p(x,t) = \left(1 + \frac{B}{2A}\right) \frac{P^2 S}{16\pi\rho c^4 \alpha x} \frac{\partial^2}{\partial t^2} \left[f\left(t - \frac{x}{c}\right)\right]^2$$

where $P$ is the pressure amplitude of the primary signal, $S$ is the surface area of the transducer, $f(t-x/c)$ is the envelop of the primary signal, $x$ is the propagation distance, $t$ is the time, $B/A$ the nonlinear parameter of the medium, $\rho$ is the density, $c$ the sound speed and $\alpha$ is the absorption coefficient.

Our first studies were done in the lab using planar transducers. This work concluded with positive results regarding the feasibility of the bipolar pulse generation with the parametric technique [9].

The next step was to deal with the cylindrical symmetry. A new setup was configured in order to, on the one hand, study in more detail the influence of the signal envelope used in emission with respect to the secondary beam generated in the medium and, on the other hand, reproducing the desired 'pancake' pattern of emission. For this purpose a single cylindrical transducer (Free Flooded Ring SX83, (FFR-SX83), manufactured by Sensor Technology Ltd., Canada) was used. This work is described in [10] and the results obtained agreed with the expectations from parametric theory and previous measurements, but now being able to generate the signals for almost cylindrical propagation. Afterwards, new measurements over longer distances were made in a larger pool using the same equipment and conditions [11]. The main objective was to compare the pressure level of both beams, primary and secondary ones, as a function of the distance since the difference of the attenuation factors is clear evidence that confirms the parametric acoustic generation of the secondary bipolar pulse. In these measurements, the obtaining of similar directivity patterns for both beams was also confirmed. An opening angle with a FWHM of about 14 ° was observed.

Following these results, the parametric acoustic sources technique can be considered a good tool to develop a transmitter able to mimic the acoustic signature of a UHE neutrino interaction. Moreover, it presents the advantage of a compact design with respect to other classical solutions [12]. To obtain directive beams with a linear phased array, the use of higher frequencies implies that fewer elements with smaller spatial separation are needed. Such a device would be easier to install and deploy in an undersea neutrino telescope.

## 3. Compact acoustic system: design and tests

### 3.1 First prototype of the compact array

With the aim of reproducing the 'pancake' directivity a three element array configuration solution is proposed. Moreover, the array will improve the signal level of the non-linear beam generated at the medium to cover long distances. The array is composed of three FFR-SX83 transducers. Each one has a diameter of 11.5 cm and is 5 cm high. The transducers usually work in the 5-20 kHz range, but for our application a second resonance peak at about 400 kHz is used, which is the frequency used for the primary beam. The array developed for the tests has a separation between elements of 2 cm, i.e. the active part of the array has a total height of 20 cm. The three elements are fixed by three bars with mechanical supports to maintain the linear array configuration. Measurements for the array characterization have been made in a pool of 6.3 m length, 3.6 m width and 1.5 m depth. The array is used as emitter, and the receiving hydrophone used to measure the acoustic waveforms is a spherical omnidirectional transducer (model ITC-1042) connected to a 20 dB gain preamplifier (Reson CCA 1000). A DAQ system is used for emission and reception. To drive the emission, a 14 bits arbitrary waveform generator (National Instruments, PCI-5412) has been used with a sampling frequency of 10 MHz. This feeds a linear RF amplifier (1040L, 400W, +55 dB, Electronics & Innovation Ltd.) used to amplify the emitted signal. For the reception, an 8 bit digitizer (National Instruments, PCI-5102) has been used with a sampling frequency of 20 MHz. The recorded data are processed and a band-pass filter (FIR, corner frequencies 5 kHz and 100 kHz) is applied to extract the secondary beam signal, as well as the relevant signal parameters (amplitude, shape and duration) to study.

An example of a received signal and the primary and secondary beams obtained after applying the band-pass filter used are shown on Figure 1 (the secondary beam has been amplified by a factor 3 for a better visibility). Figure 2 shows the directivity pattern measured along a line parallel to the axis of the array at a distance of 2.7 m. It allows studying the angle distribution of both beams. Instead of having for the primary beam a thin peak, a wide peak with no clear maximum is observed. This is due to the fact that the measurements were done at a distance which cannot be considered very large, and therefore the signals from the different transducers are not totally synchronous at 0°. Moreover, some of the variations of the primary beam may be due as well to the interference (constructive or destructive) of the individual beams, which are not only observed from the variations point to point, but even in the same point. This might be the reason for the large differences observed in the statistical uncertainties of the primary beam measures. Despite the small distance, the behaviour of the secondary beam is smoother and the FWHM measured with the array is about 7° (σ=3°), that is, smaller than for the primary beam, and sensitively smaller than for a single element where the FWHM was about 14° (σ =6°). Moreover, considering that the signals for the three elements will be better synchronized at large distances (larger than 100 m), it seems possible to achieve a 'pancake' directivity with an aperture of the 1° order (σ ~1°) for large distances, as expected from our calculations [11].

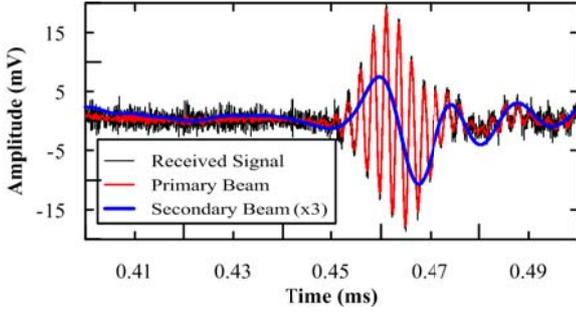

Figure 1. Example of a received signal. Primary and secondary beams obtained after applying the band-pass filter (the secondary beam has been amplified by a factor 3 for a better visibility).

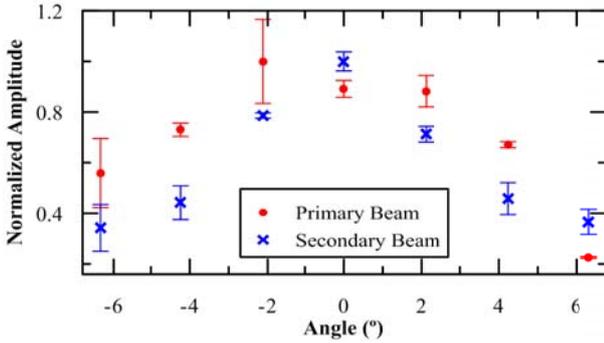

Figure 2. Directivity patterns of primary and secondary beam measured with the array.

### 3.2 Propagation to large distances

In order to show the effect of the propagation caused on the bipolar parametric signal, signals recorded experimentally have been propagated to a distance of 1 km. An algorithm which works in the frequency domain corrects the frequency dependence of the hydrophone sensitivity and propagates each spectral component considering the geometrical attenuation of the pressure beams as 1/r and the absorption coefficient [13] calculated for the sea conditions in the ANTARES site is used. Figure 3 shows a signal propagated to 1 km (original signal was measured in the lab with a distance of 0.7 m). In this case, no additional filter is applied, the propagation medium acts as a natural filter. To be exact, there is still a small high-frequency component which is not observed at distances of 1.2 km (or higher). Notice that the high-frequency signal was three orders of magnitude higher than the secondary beam at ~1 m distance. It appears as well a kind of DC offset, it is due to the very low-frequency components of the signal (probably 50 Hz) which are also propagated. As conclusion, using a single element it is expected to have a bipolar pulse with 25 mPa peak-to-peak amplitude at 1 km distance from the sensor and 0° direction. Considering the array configuration with three elements feed in phase with the maximum power, which is not possible with the current electronics, it is expected to have, at least, 0.1 Pa peak-to-peak, which is a value clearly above the AMADEUS threshold (20 mPa) [5], and corresponds to the pressure reference of the order of a neutrino interaction of $10^{20}$ eV at 1 km distance.

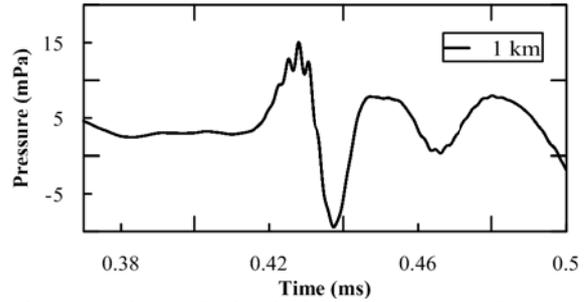

Figure 3. Signal obtained after 1 km propagation.

### 3.3 Second prototype of the compact array

The final goal is to have a system able to carry out several tasks related to acoustics in an underwater neutrino telescope using the same transmitter: acoustic detection calibration, calibration of the acoustic receivers sensitivity, and signals emission for positioning tasks. Certainly, this could reduce the cost and facilitate the deployment and operation at deep sea.

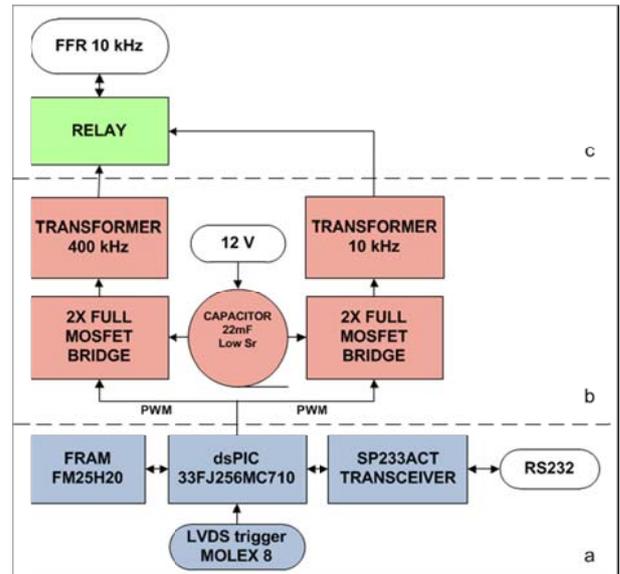

Figure 4. Electronics block diagram: Communication and control (a), emission (b) and commuter between the two operation modes (c).

### 3.3.1 Mechanics

For the second prototype the transducers are fixed on a rigid axis using flexible polyurethane (EL110H, Robnor Resins Ltd). Due to the nature of the cured polymer, this offers water resistance and electrical insulation for high frequency and high voltage applications. The resulting compact design has a length of 17.5 cm for the active region. In addition, a mechanical structure has been built for handling the array from a boat and controlling its rotation angle (orientation of the emission). This is a critical aspect to be able to point to the receivers due to the high directivity.

### 3.3.2 Associated electronics

Developments in electronics have been made with the aim of having an autonomous and optimized compact system able to work in different frequency ranges for different applications, see Figure 4. An electronic device that controls the transmitter, generates and amplifies the signals in order to have enough acoustic power to achieve the emission in nonlinear regime. This is an essential requirement for parametric generation and it is necessary for calibration and/or positioning purposes in order to be detected from distant acoustic receivers.

A novelty in the design is the use of Pulse Width Modulation technique (PWM) [14]. With this, it is possible to emit arbitrary intense short signals, and therefore to emit the necessary 'modulated signal' with the goal of obtaining a secondary beam with the desired specifications. This technique has been already implemented for the acoustic transceiver electronics of the positioning systems for the future KM3NeT neutrino telescope [15-17].

Some of the advantages this technique offers are:
- Efficiency: System uses class D amplification. Transistors are working on switching mode, suffering less power dissipation.
- Simplicity of design: Analog-digital converters are not needed. It is possible to feed directly the amplifier with the digital signal modulated by the PWM technique.
- Lightweight and compact system: It is not necessary to install large heat sinks at the amplifier transistors.
- Minimum power consumption in stand-by mode: this is required for a system of a deep-sea detector. Moreover, it allows storing the energy in the amplifier capacitor very fast and efficiently for using it in the following emission.

### 4. Conclusions and future steps

Considering all the results obtained in relation to the studies of parametric acoustic sources and to the first array prototype it seems clear that the solution proposed based on parametric acoustic sources could be considered a good candidate to generate the acoustic neutrino-like signals, with both specific characteristics of the predicted signal: bipolar shape in time and 'pancake directivity'.

Moreover, due to the transmitter system versatility, the array plus the electronics described for the last prototype could be implemented to carry out several tasks related to acoustic emission for underwater neutrino telescopes: positioning, acoustic detection calibration and calibration of the receiver hydrophone sensitivities. This may simplify the set of acoustic transmitters needed in a neutrino telescope, and thus may reduce the complexity of the system and the costs with respect to the use of multiple devices.

The future work will consist in completing the compact transmitter prototype (array+electronics) and characterizing it in the lab and in situ. For the last part, tests of the transmitter using the AMADEUS system are foreseen, either using the array from a vessel in a sea campaign or maybe integrated in the ANTARES infrastructure.

### Acknowledgements

This work has been supported by the Ministerio de Ciencia e Innovación (Spanish Government), project references FPA2009-13983-C02-02, ACI2009-1067, Consolider-Ingenio Multidark (CSD2009-00064). It has also being funded by Generalitat Valenciana, Prometeo/2009/26.